\documentclass{iaus}
\usepackage{graphicx}
\usepackage{natbib}

\title[Magneto-hydrodynamic waves in roAp stars] 
{Simulations of magneto-hydrodynamic waves in atmospheres of
roAp stars}

\author[Khomenko and Kochukhov]   
{E. Khomenko$^{1,2}$ \and O. Kochukhov$^3$}

\affiliation{$^1$Instituto de Astrof\'{\i}sica de Canarias, 38205,
C/ V\'{\i}a L{\'a}ctea, s/n, Tenerife, Spain \break $^2$Main
Astronomical Observatory, NAS, 03680, Kyiv, Ukraine \break
 $^3$Department of Physics and Astronomy, Uppsala
University, Box 515, SE-751 20, Sweden \break  email:
khomenko@iac.es}

\pubyear{2004}
\volume{xxx}  
\pagerange{119--126}
\date{?? and in revised form ??}
\jname{Proceedings Title IAU Symposium}
\editors{A.C. Editor, B.D. Editor \& C.E. Editor, eds.}
\begin{document}

\maketitle

\begin{abstract}
We report 2D time-dependent non-linear magneto-hydrodynamical
simulations of waves in the atmospheres of roAp stars. We explore
a grid of simulations in a wide parameter space. The aim of our
study is to understand the influence of the atmosphere and the
magnetic field on the propagation and reflection properties of
magneto-acoustic waves, formation of shocks and node layers.
\keywords{MHD; stars: magnetic fields; stars: oscillations}
\end{abstract}


RoAp stars are late-A, chemically peculiar stars with effective
temperatures between 6500--8100 K and global dipolar-like magnetic
fields with strengths between 1--25 kG. They show low-order
angular degree $p$-mode pulsations with periods between 4 and 22
minutes. These non-radial pulsations are aligned with dipolar
field axis. Magneto-acoustic oscillations in peculiar A stars are
of particular interest due to unique opportunities to study the
interaction of pulsations, chemical inhomogeneities, and strong
magnetic fields. Recent reviews on the properties of these stars
and their pulsations can be found in \citet{Kurtz2008, Cunha2007,
Kochukhov2007, Kochukhov2008}.

We solve the basic 2D non-linear adiabatic equations of the ideal
MHD by means of the numerical code described by
\citet{Khomenko+Collados2006, Khomenko+Collados2007}. We assume
that: (1) the magnetic field varies on spatial scales much larger
than the typical wavelength, allowing the problem to be solved
locally for a plane-parallel atmosphere with a homogeneous
inclined magnetic field; (2) waves in the atmosphere are excited
by low-degree pulsation modes with radial velocities exceeding
horizontal velocities. The unperturbed atmospheric model has an
effective temperature of $T_{\rm eff}$ = 7750 K and gravitational
acceleration at the surface log$_{10}g=4.0$. The simulation grid
covers the magnetic field strength $B=1 - 7$ kG; magnetic field
inclination to the local vertical $\gamma = 0 - 60$ degrees and
driving periods $T= 6-13$ minutes.
An example of the wave pattern developed in the simulations is
given in Fig. 1. We use the longitudinal and transversal field
projections of the velocity in order to separate clearly the fast
and slow MHD waves. Our first results pick up some observed
properties of the roAp stars pulsations, such as: rapid growth of
the wave amplitude with height, presence of the node surfaces,
large variations in pulsation properties depending on the
parameters of the model. Velocity signal observed in the upper
atmospheric layers of roAp stars is mostly due to running slow
mode (acoustic) waves propagating along the inclined magnetic
field lines. The node structures and the rapid phase variations at
the lower atmospheric layers are due to multiple reflections and
interference of the slow and fast mode waves.
The disc-integrated velocity signal produced by the atmospheric
pulsations of such a star will depend in a complex way on the
inclination of the magnetic axis with respect to the observational
line of sight and will be a subject of our further study.

\begin{figure}
\center
 \includegraphics[width=11cm]{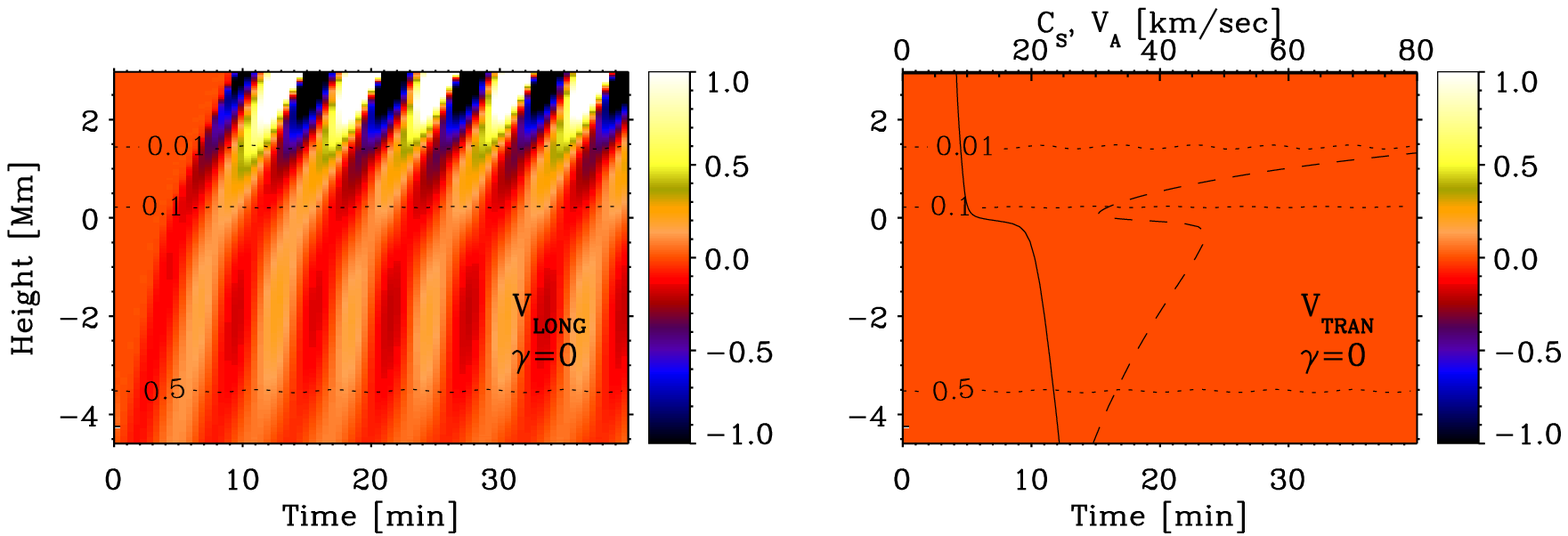}
 \includegraphics[width=11cm]{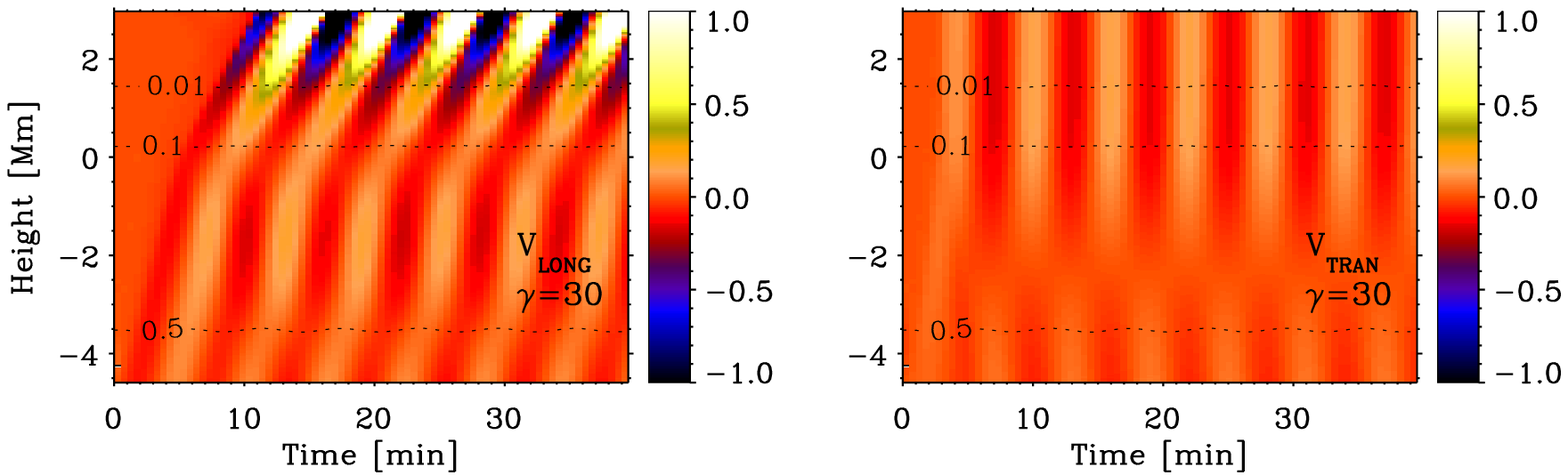}
 \includegraphics[width=11cm]{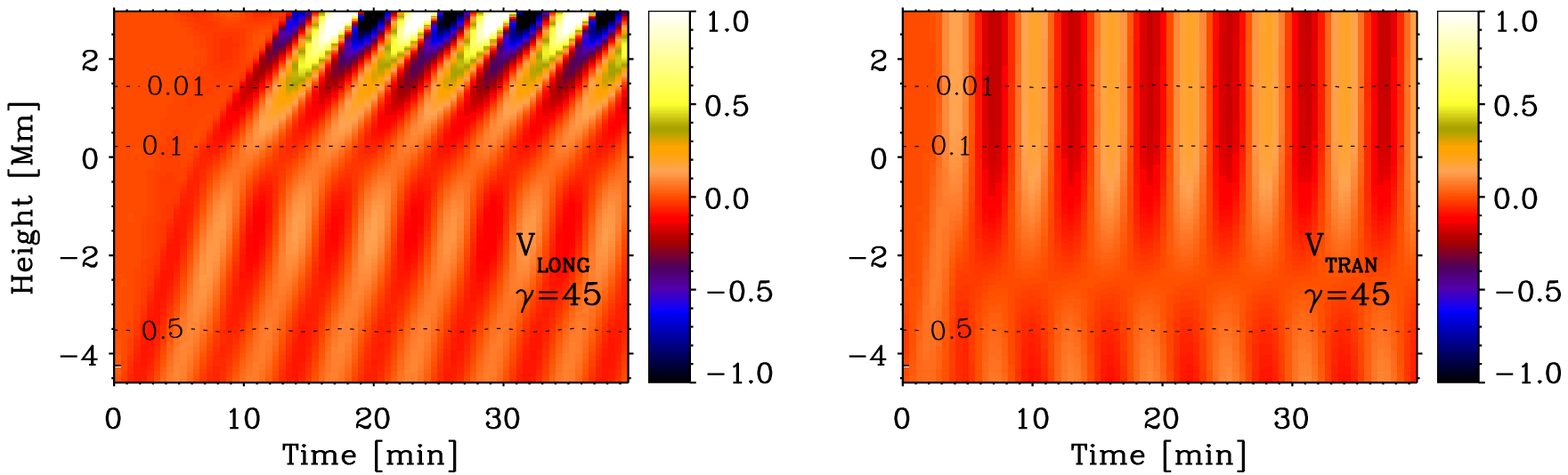}
 \includegraphics[width=11cm]{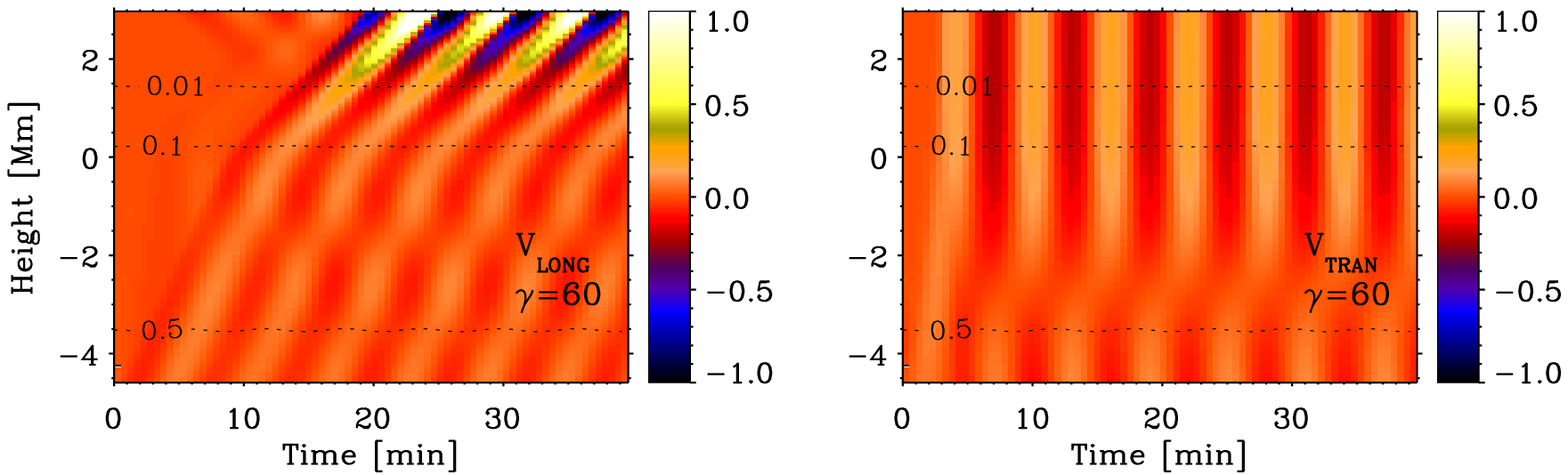}
  \caption{Height-time variations of the longitudinal (parallel to the field, left)
  and transversal (perpendicular to the field, right) velocities for B=1 kG, T=360 sec
  at latitudes where the inclination equals 0, 30, 45 and 60  degrees. The color bars
  give velocity   scale in km/sec. Zero height corresponds to the photospheric base.
  Dotted lines marked with numbers are contours of constant $c_S^2/v_A^2$ . Height
  dependences of the sound speed $c_S$ (solid line) and Alfv\'en speed $v_A$
  (dashed line) are plotted over the top panel, the scale is given by the upper axis.}\label{fig:wave}
\end{figure}
%



\end{document}